\documentclass[prd,aps,twocolumn,nofootinbib]{revtex4}
\usepackage{natbib}
\usepackage{graphicx}
\usepackage{color}
\usepackage{hyperref}
\usepackage{siunitx}
\def\gr{$\gamma$-ray}

\begin{document}
\title{Detectability of large correlation length inflationary   magnetic field with Cherenkov telescopes }
\author{
Alexander Korochkin$^{1,2,3}$,
Andrii Neronov$^{1,4}$,  Guilhem Lavaux$^{5}$, 
Marius Rams\o y$^{1,5}$ and 
Dmitri Semikoz$^{1,2,6}$ }
\affiliation{$^1$ Université de Paris, CNRS, Astroparticule et Cosmologie,  F-75006 Paris, France\\
$^2$ Institute for Nuclear Research of the Russian Academy of Sciences, 60th October Anniversary Prospect 7a, Moscow 117312, Russia\\
$^3$ Novosibirsk State University, Pirogova 2, Novosibirsk, 630090 Russia\\
$^4$Laboratory of Astrophysics, Ecole Polytechnique Federale de Lausanne, 1015, Lausanne, Switzerland \\
$^5$Institut d’Astrophysique de Paris (IAP), CNRS \& Sorbonne Universit\'e, UMR 7095, 98 bis bd Arago, F-75014 Paris, France\\
$^6$National Research Nuclear University MEPHI (Moscow Engineering Physics Institute),
Kashirskoe highway 31, 115409 Moscow, Russia}

\begin{abstract}
Magnetic fields occupying the voids of the large scale structure may be a relic from the Early Universe originating from either Inflation or from cosmological phase transitions. We explore the possibility of identifying the inflationary origin of the void magnetic fields and measuring its parameters with $\gamma$-ray astronomy methods.  The large correlation length inflationary field is expected to impose a characteristic asymmetry of extended \gr\ emission that is correlated between different sources on the sky. We show that a set of nearby blazars for which the extended emission is observable in the 0.1-1 TeV band with CTA can be used for the test of inflationary origin of the void magnetic fields. \end{abstract}
\maketitle
\section{Introduction}

Observations of extended and delayed gamma-ray emission around extragalactic sources of TeV \gr s provides a possibility of measurement of magnetic field in the voids of the Large Scale Structure (LSS) \cite{plaga95,Neronov:2007zz,Neronov:2009gh}. This emission is generated by electron-positron pairs deposited by the pair production by \gr s on the Extragalactic Background Light. The combination of data from current generation Cherenkov telescopes, HESS, MAGIC and VERITAS with data from the Fermi-LAT telescope currently constrains the void field strength to be stronger than $\sim 10^{-17}$~G \cite{Neronov:1900zz,Taylor:2011bn,Biteau:2018tmv}. The next-generation Cherenkov Telescope Array (CTA) will provide a possibility to explore the magnetic field over a wide range of possible strength and correlation lengths, up to the field strength of the order of $B\sim 10^{-11}$~G  \cite{Korochkin:2020pvg,Vovk:2021aqb}. 

Several physical phenomena that took place a fraction of a second after the Big Bang can be responsible for the generation of a relic magnetic field. First order phase transitions that might have happened at quark confinement or during the Electroweak epoch can produce short correlation length magnetic field that evolves through turbulent decay toward a magnetic field configuration with correlation length and strength satisfying a relation  $
\lambda_B\sim 0.1 [B/10^{-12}\mbox{ G}]$~kpc  today \cite{Banerjee:2004df,Kahniashvili:2012uj}.  Alternatively, a field generated at the epoch of inflation can have a very large correlation length, up to the present day Hubble scale \cite{1992ApJ...391L...1R,Garretson:1992vt,Gasperini:1995dh,Giovannini:2000dj,Demozzi:2009fu,Ferreira:2013sqa,Durrer:2013pga}.

Magnetic fields from galactic outflows, if they efficiently pollute the voids (as suggested e.g. by \citet{2006MNRAS.370..319B}), would be distinguishable from both inflationary and phase transition field based on their galaxy scale correlation length (10-100 kpc).  This is much larger than that of the phase transition induced field, but not reaching the Hubble scale of the correlation length of the inflationary field. However even though magnetised outflows from galaxies can spread non-cosmological magnetic fields in the intergalactic medium, these outflows are most likely not strong enough to fill the voids \cite{marinacci18,Bondarenko:2021fnn}. This suggests that the volume-filling magnetic field in the voids is a relic from the Early Universe \cite{Durrer:2013pga}.

In what follows we explore the possibility of distinguishing between these two possibilities  observationally. Large correlation length fields break isotropy by selecting a unique direction in a cosmologically large volume. This selected direction imposes a correlated asymmetry on magnetic field dependent extended emission patterns around \gr\ sources across the sky. We explore if it is possible to detect this asymmetry and measure the inflationary magnetic field direction.  We use a magnetic field generated from a realistic model of the LSS derived from {\sc Borg} constrained cosmological simulations  \cite{2013MNRAS.432..894J} that reproduces the location of known LSS elements (galaxies, clusters) in the local Universe. Our analysis relies on magneto-hydrodynamic (MHD) simulations using the {\sc ramses} code \cite{2002A&A...385..337T} to estimate the effect of the structure formation on the initial magnetic field configuration.  Modelling the properties of secondary \gr\ signal is performed with the {\sc CRbeam} code \cite{Berezinsky:2016feh}. We use calculations of electromagnetic cascades along lines of sight to known nearby blazars to estimate the influence of large correlation length magnetic field on the properties of secondary \gr\ signal from the cascade.

%%%%%%%%%%%%%%%%%%%%%%%%%%%%%%%%%%%%%%%%%%%%%%%%%
\section{Modelling of the magnetic field}
%%%%%%%%%%%%%%%%%%%%%%%%%%%%%%%%%%%%%%%%%%%%%%%%%

%We calculate CMF within the distance up to 200~Mpc around the Milky Way by running MHD simulations using the {\sc Ramses-MHD} code \citep{ramses-MHD} as a post-processing of constrained cosmological simulations that reproduce positions of real galaxies and galaxy clusters 2M++ survey \citep{2M++}. These simulations use the algorithm dubbed `Bayesian Origin Reconstruction from Galaxies', or {\sc Borg} \citep{borg}.

We calculate intergalactic magnetic field (IGMF) out to a distance of 200~Mpc from the Milky Way by running MHD simulations using the {\sc Ramses-MHD} code \citep{ramses-MHD} on the initial conditions (ICs) from {\sc Borg} \citep{Borg}.  Using a Markov Chain Monte Carlo approach, the {\sc Borg} methodology generates ICs that are constrained to reproduce the structure of positions of real galaxies and galaxy clusters of the 2M++ survey \citep{2M++} within 200 Mpc cube around Milky Way. 
A cosmological zoom of one variant of the {\sc Borg} ICs was produced using {\sc Music} algorithm  \citep{HahnMusic} to generate a high resolution region 200~Mpc around the Milky Way with spatial resolution at $z=0$ of $0.7 h^{-1}$~Mpc and mass resolution $\rm \num{2.1e10} h^{-1} M_{\odot}$.  This is a factor 4 increase over the original {\sc Borg} ICs spatial resolution.  Note however that the density fluctuations on this scale are unconstrained.  The initial configuration of the magnetic field is uniform across the simulation volume with strength up to $10^{-12}$~G. Its direction is chosen to be in the direction of north (${\it Dec}=90^\circ$) of the ICRS coordinate system. The hydrodynamical variables and magnetic field were evolved on the AMR grid using the HLLD solver \citep{Miyoshi2005} with the MinMod slope limiter being used to reconstruct variables at cell interfaces.

The simulation was run without cooling, star formation or feedback as the focus of this study are cosmological voids where the impact of processes are likely to be small \cite{Bondarenko:2020moh,Bondarenko:2021fnn}. As a result of this the magnetic field is a smoothed tracer of the dark matter density field.  Only sufficiently massive objects are capable of producing large enhancements of the gas density field.  Regions with overdensities greater than approximately 3 were thought to be close enough to structures that generate magnetized outflows that they are at risk of being  ``polluted" by magnetic fields of non-cosmological origin, as indicated in \cite{marinacci18}. The magnetic field strength in such regions may deviate from simple scaling imposed by the amplification  due to pure adiabatic contraction.  As a result we therefore mark as high magnetic field region any area with an overdensity greater than 3 averaged over a thin cylinder with radius 2.7~Mpc (3 cells in the zoom region) along the line of sight during the course of our analysis.

{\sc Ramses-MHD} was used due to its implementation of a constrained transport scheme \citep{MHD-CT}.  Potentially spurious amplification of the magnetic field seen in codes that use divergence cleaning schemes \cite[e.g.][]{Stasyszyn2015,Mocz2016} are suppressed by this scheme. In any case, this effect is most important in high density regions such as the centres of clusters and should have a minimal effect in this work.

%%%%%%%%%%%%%%%%%%%%%%%%%%%%%%%%%%%%%%%%%%%%%%%%%
\section{Nearest TeV blazars}
%%%%%%%%%%%%%%%%%%%%%%%%%%%%%%%%%%%%%%%%%%%%%%%%%

%%%%%%%%%%%%%%%%%%%%%%%%%%%%%%%%%%%%%%%%%%%%%%%%%%%%
\begin{table}[]
    \begin{tabular}{|c|c|c|c|c|}
    \hline
    Name & RA & Dec & $z$ & $F_{1\mathrm{TeV}}$, TeV cm$^{-2}$ s$^{-1}$  \\
    \hline
        Mkn 421& 166.11 & 38.21 & 0.031 & $2\times 10^{-11}$\\
        Mkn 501 & 253.47 &39.76	&0.033 & $1\times 10^{-11}$\\
        QSO B2344+514 & 356.77	& 51.7 &0.044 & $4\times 10^{-12}$\\
        Mkn 180 & 174.11 & 70.16 & 0.046 & $8\times 10^{-13}$\\
        1ES 1959+650 & 299.99 &	65.15 &0.047 & $6\times 10^{-12}$\\
       AP Librae & 229.42 &	-24.37 & 0.04903 & $4\times 10^{-13}$\\
        TXS 0210+515 & 33.57&	51.75 & 0.04913 & $2\times 10^{-13}$\\
    \hline
    \end{tabular}
    \caption{List of nearby TeV blazars considered in the study. }
    \label{tab:list}
\end{table}
%%%%%%%%%%%%%%%%%%%%%%%%%%%%%%%%%%%%%%%%%%%%%%%%%%%%

There are several known TeV blazars whose position falls within the high resolution simulation volume.  The source list is given in Table \ref{tab:list}.  All the sources are well-established TeV emitters. The proximity of the sources enables measurements of their intrinsic spectra attenuated by the pair production effect up to 10~TeV. This is important because the extended emission in the energy range above 100~GeV is produced by electrons and positrons injected in interactions of \gr s with energies above 10~TeV. Measurement of the flux above 10~TeV allows us to obtain reliable estimates of the expected power of the secondary flux in the $E>100$~GeV range. 

The lines of sight toward the sources listed in Table \ref{tab:list} are oriented at different angles with respect to the direction of initial homogeneous field chosen in simulation.  This provides a possibility to study how variations of the misalignment of the cosmological field direction affects the observational appearance of the extended emission. Lines of sight toward two sources, 1ES 1959+650 and Mkn 180 are more strongly aligned to the direction of the magnetic field than, say, the lines of sight to Mkn 421 and Mkn 501. The line of sight toward AP Librae has the largest misalignment angle in our sample. 

In the following sections we show results for the three brightest sources in our sample Mkn~501, Mkn~421 and 1ES~1959+650, see Table \ref{tab:list}.

%%%%%%%%%%%%%%%%%%%%%%%%%%%%%%%%%%%%%%%%%%%%%%%%%
\section{Structure of magnetic fields along the lines of sight}
\begin{figure}
    \includegraphics[width=0.9\linewidth]{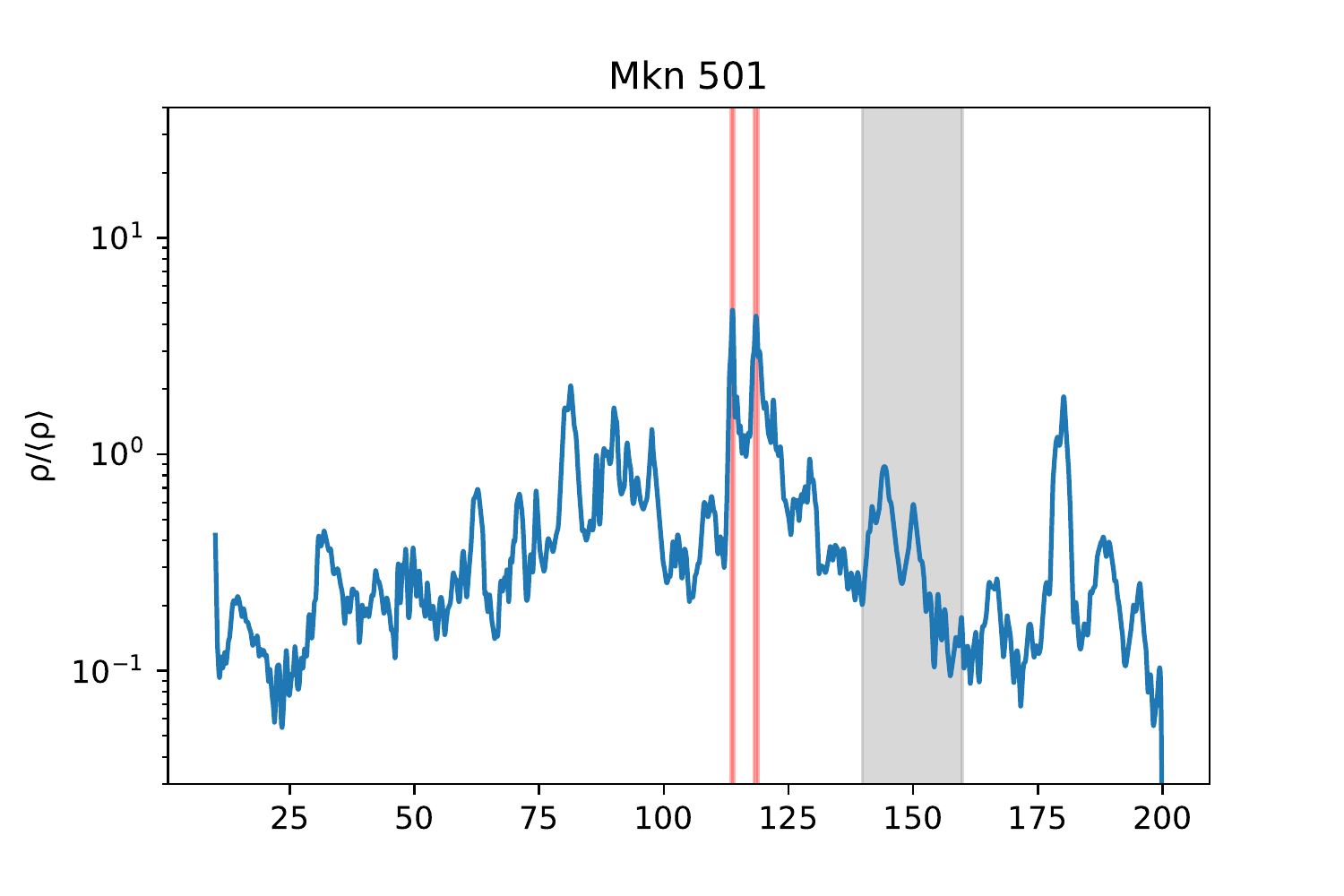}
    \includegraphics[width=0.9\linewidth]{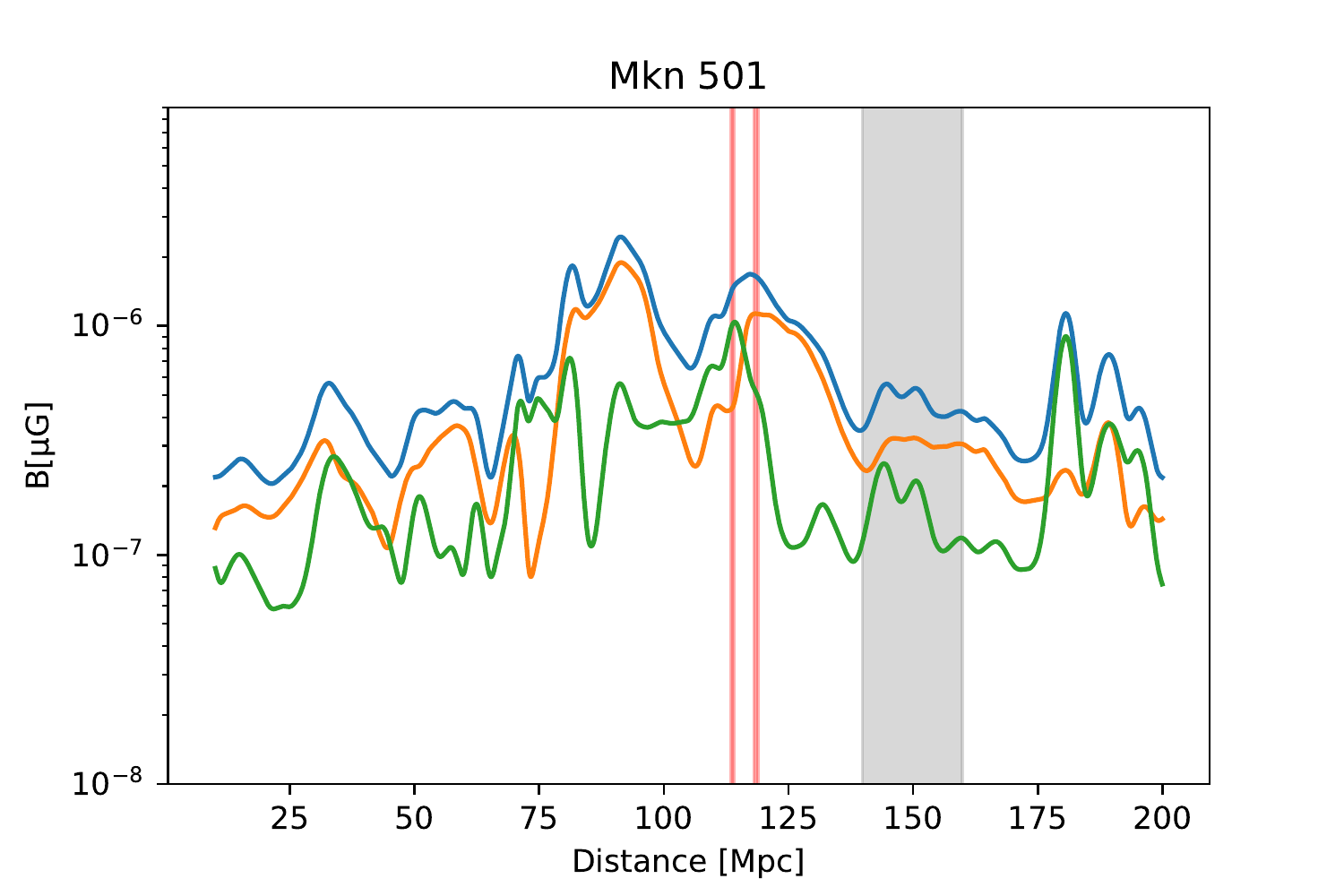}
    \includegraphics[width=0.9\linewidth]{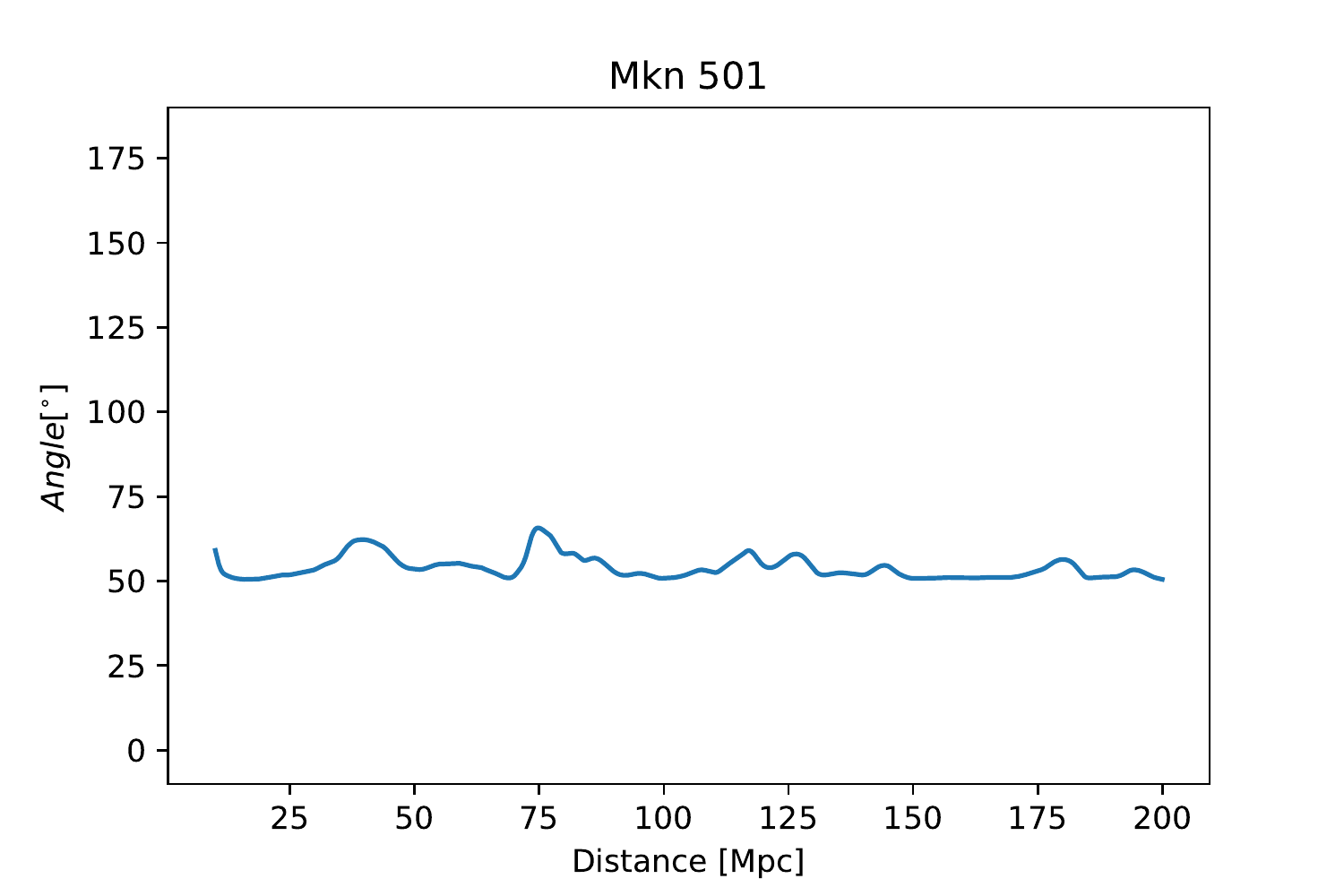}
    \caption{Density profile(top), magnetic field (middle) and angle between the magnetic field and the primordial direction (bottom) for a line of sight in the direction of Mkn 501. The magnetic field is split into total (blue), along the line of sight (orange) and perpendicular to line of sight (green).  The location of the source with uncertainty is shown with a grey band.  
 Regions where the overdensity is greater than 3 are indicated by red bands.}
    \label{fig:Mkn501}
\end{figure}

\begin{figure}
    \includegraphics[width=0.9\linewidth]{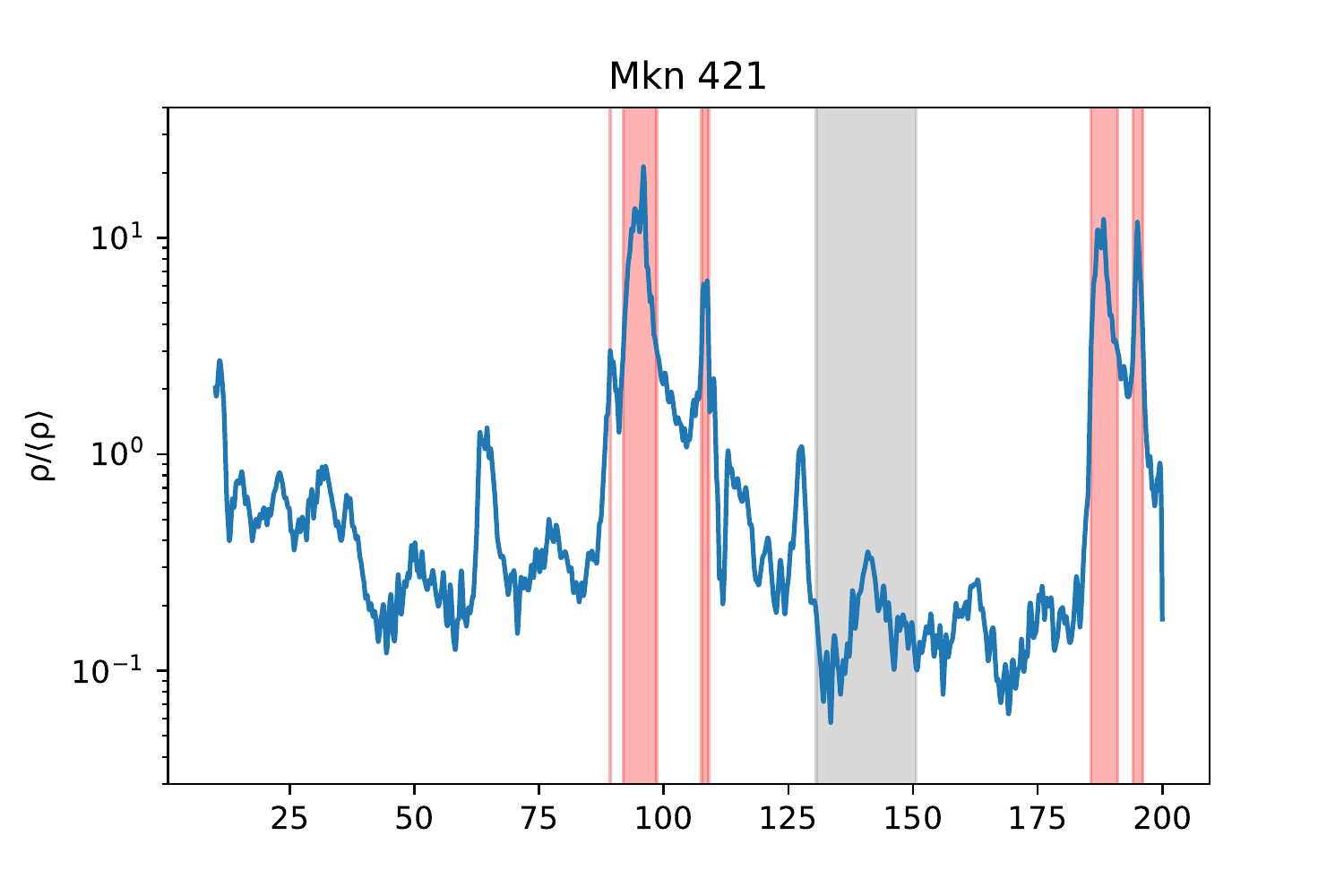}
    \includegraphics[width=0.9\linewidth]{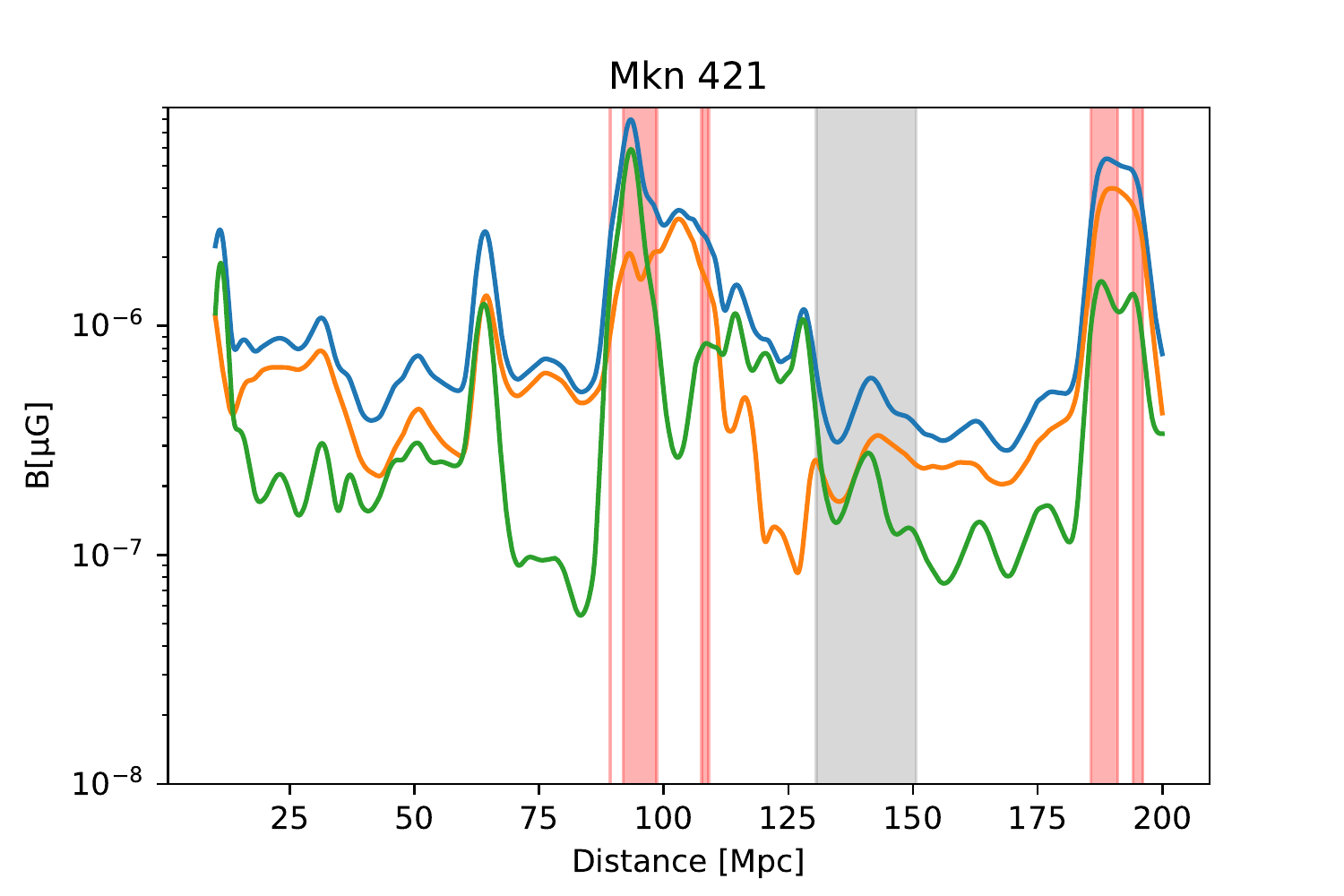}
    \includegraphics[width=0.9\linewidth]{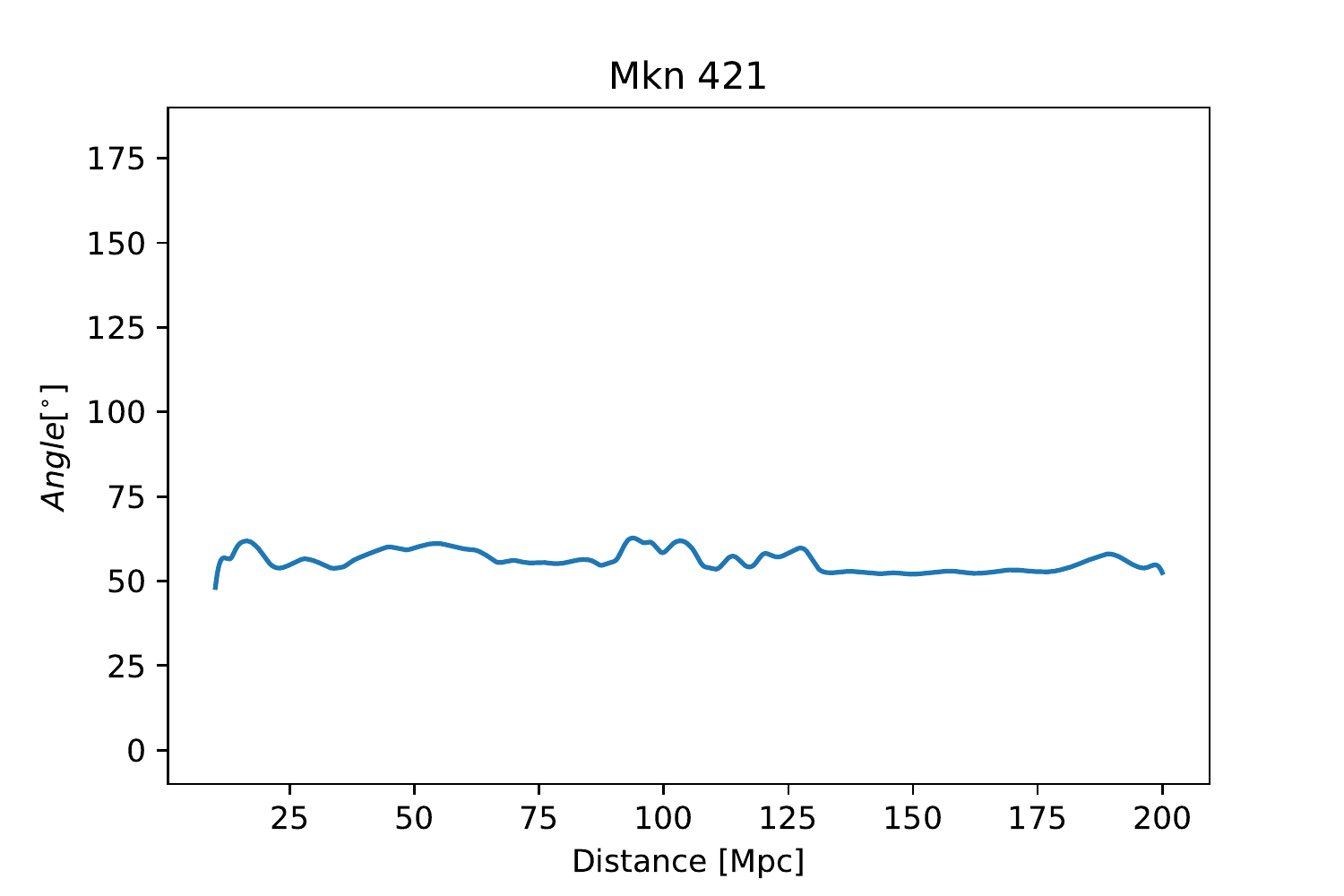}
    \caption{Same as Fig. \ref{fig:Mkn501} for Mkn 421.}
    \label{fig:Mkn421}
\end{figure}

\begin{figure}
    \includegraphics[width=0.9\linewidth]{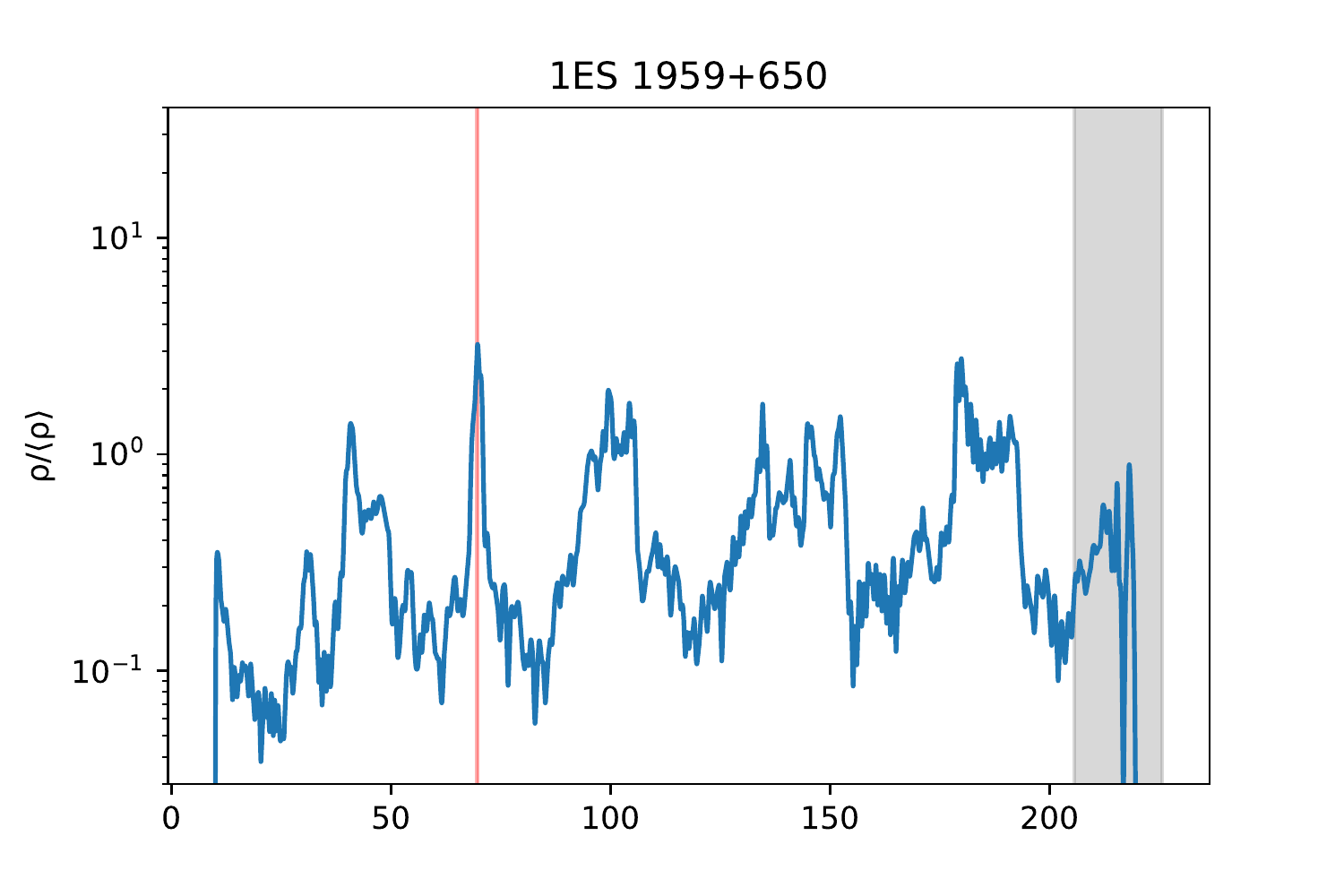}
    \includegraphics[width=0.9\linewidth]{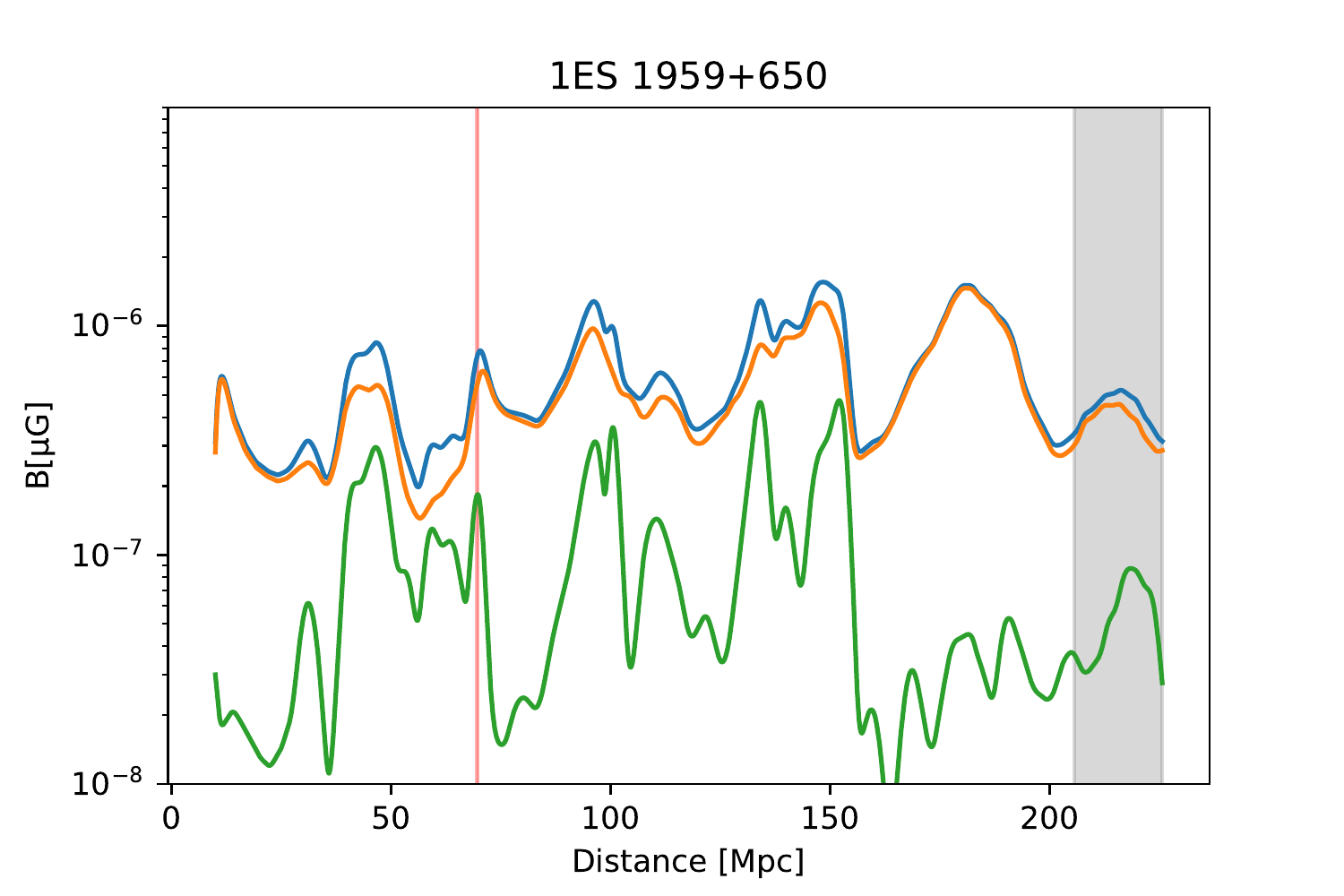}
    \includegraphics[width=0.9\linewidth]{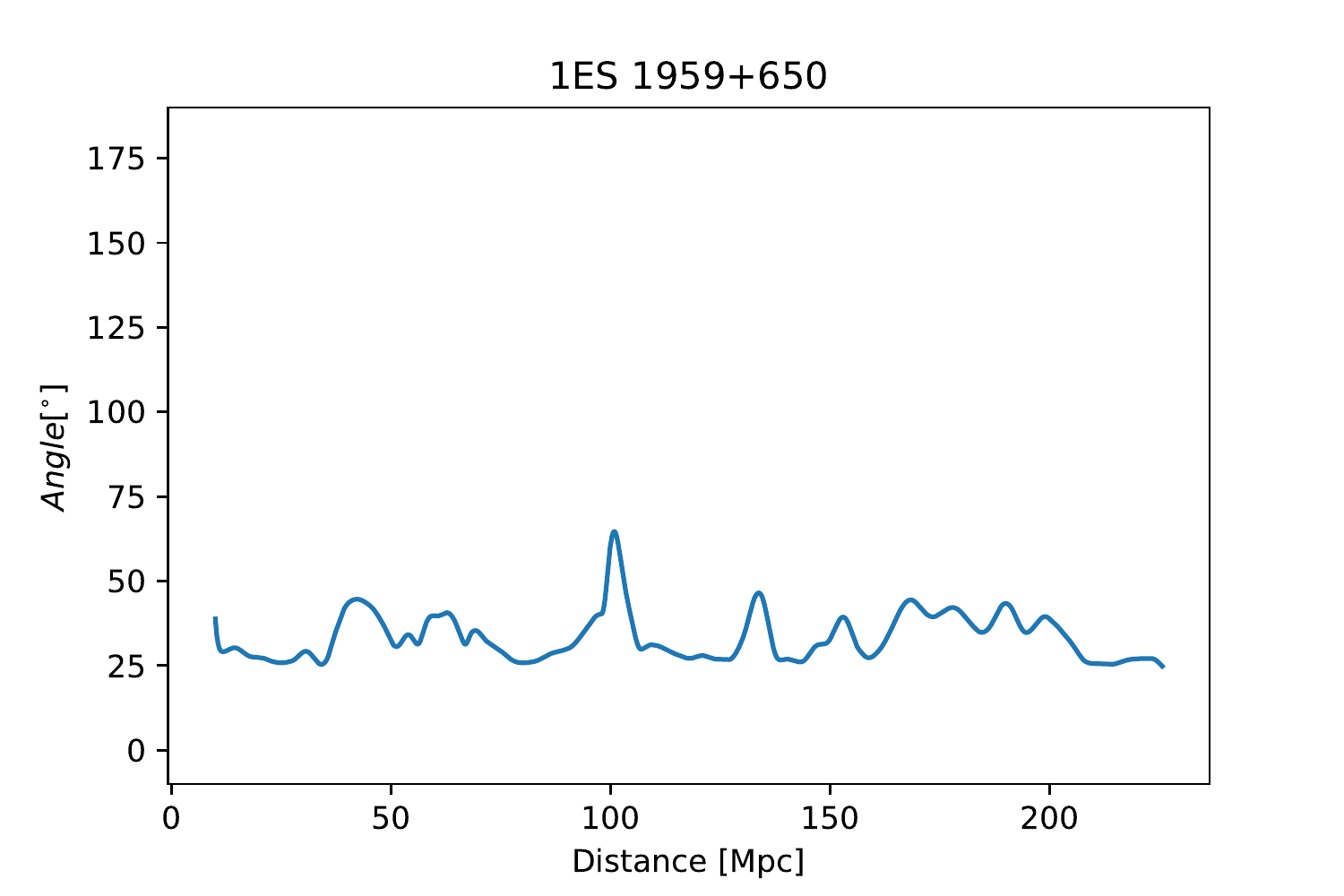}
    \caption{Same as Fig. \ref{fig:Mkn501} for $1ES 1959+650$.}
    \label{fig:1ES_1959+650}
\end{figure}

%%%%%%%%%%%%%%%%%%%%%%%%%%%%%%%%%%%%%%%%%%%%%%%%%
Figs. \ref{fig:Mkn501}, \ref{fig:Mkn421} and \ref{fig:1ES_1959+650} show the line-of-sight profiles of magnetic field toward Mkn 501, Mkn 421 and 1ES 1959+650. The top panels of the figures show the line-of-sight profiles of the density field. Regions with overdensities above $3$ that are excluded from our further analysis because of possible ``contamination" of the magnetic field by baryonic feedback \cite{Bondarenko:2021fnn}
are indicated with vertical red bands. In those regions secondary electrons are randomized by high magnetic fields, they emit secondary photons in random directions and their signal in direction to observer can be safely neglected. Grey bands indicate the distance to the blazars. For all three examples, the \gr\ sources are situated in moderately overdense regions that are unlikely to be affected by strong magnetised outflows from galaxies. None of the three sources 
  have a line-of-sight aligned with a filament of the LSS. This makes all the three sources suitable for the IGMF measurement. 

The middle and bottom panels of the figures show the strength and orientation of the IGMF along the lines of sight. One can see clear differences of the IGMF structure for the three sources. The component of the magnetic field perpendicular to the line of sight is strongly suppressed in the case of 1ES 1959+650, because of the alignment of the cosmological field direction with the line of sight.

%%%%%%%%%%%%%%%%%%%%%%%%%%Mkn 501 %%%%%%%%%%%%%%%%%%%%%%%%% 

%%%%%%%%%%%%%%%%%%%%%%%%

The lower panels of the figure show the position angle of the perpendicular to the line-of-sight field component, counted from the direction toward North. Deviations from the straight horizontal lines in these panels are due to the effect of structure formation. One can see that the effect is moderate and the initial cosmological field direction is mostly preserved in the course of structure formation.  

%%%%%%%%%%%%%%%%%%%%%%%%%%Mkn 421 %%%%%%%%%%%%%%%%%%%%%%%%%

%%%%%%%%%%%%%%%%%%%%%%%%%%%%%%%%%%%%%%%%%%%%%%%%%%%

%%%%%%%%%%%%%%%%%%%%%%%%%% 1ES 1959+650 %%%%%%%%%%%%%%%%%%%%%%%%%

%%%%%%%%%%%%%%%%%%%%%%%%%%%%%%%%%%%%%%%%%%%%%%%%%%%

%%%%%%%%%%%%%%%%%%%%%%%%%%%%%%%%%%%%%%%%%%%%%%%%%%%
\section{Modelling of secondary extended gamma-ray emission}
%%%%%%%%%%%%%%%%%%%%%%%%%%%%%%%%%%%%%%%%%%%%%%%%%%%

%%%%%%%%%%%%%%%%%%%%%%%%%%%%%%%%%%%%%%%%%%%%%%%%%%%
\begin{figure*}
\includegraphics[width=\textwidth]{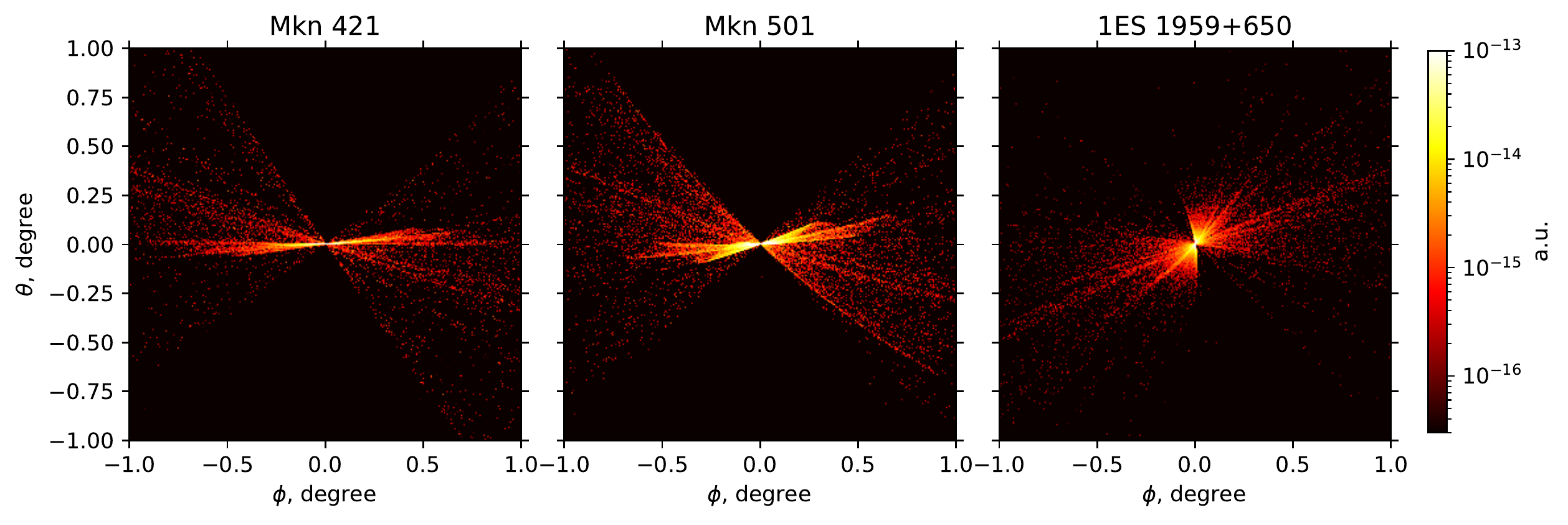}
    \caption{Images of the extended emission signal {in the energy range 200 GeV - 2 TeV} for the three brightest sources in our sample. The assumed initial cosmological magnetic field strength is $B=10^{-13}$ G. {The direction of the jet axis coincides with the direction from the source to the observer and the jet opening angle is $5^\circ$.} %\textcolor{red}{What energy range? Can you rotate the images so that the extensions become horizontal? Our magnetic field is pointing "up", to the "North", so it would be natural to see the extensions to the right and left....}
    }
    \label{fig:images}
\end{figure*}
%%%%%%%%%%%%%%%%%%%%%%%%%%%%%%%%%%%%%%%%%%%%%%%%%%%

We model the secondary cascade \gr\ signal with the {\sc CRbeam} Monte Carlo code \citep{Berezinsky:2016feh} which propagates high energy \gr s through the cosmic medium taking into account most important physical interactions: pair production by \gr\ absorption and inverse Compton scattering of secondary electrons and positrons on the EBL \citep{Franceschini08} and CMB. Electrons and positrons are also deflected in magnetic field whose strength is obtained from IGMF model described in the previous section. 

For intrinsic point source spectrum we assume broken power law spectrum with break position around 100 GeV and maximum energy 100 TeV. All primary \gr s have the same direction of initial momenta which coincides with the direction from the source to the observer. The \gr s are propagated until they reach the sphere whose center coincides with the position of the source and radius is equal to the distance to the observer. To model primary $\gamma$-ray emission into a jet with an opening angle $\alpha_\mathrm{jet}$ we select \gr s whose positions on the sphere lie inside the cone with the opening angle $\alpha_\mathrm{jet}=5^\circ$ and direction of the axis coincides with the direction of jet. We do not take into account secondary photons that were created in the regions with an overdensity above 3, see Figs. \ref{fig:Mkn501}, \ref{fig:Mkn421} and \ref{fig:1ES_1959+650}. To explore different strengths of seed magnetic field we rescale the magnetic field profile by a constant factor keeping the shape of the profile unchanged.

Figs. \ref{fig:images} shows the images of the extended emission calculated for the field strength $10^{-13}$~G for the three brightest sources in our sample. The images of extended emission around Mkn 421 and Mkn 501 reveal a clear overall azimuthal asymmetry that is imposed by the orientation of the initial homogeneous field projected on the sky. To the contrary, the image for 1ES 1959+650 appears halo-like. This is explained by the alignment of the magnetic field along the line of sight.

%%%%%%%%%%%%%%%%%%%%%%%%%%%%%%%%%%%%%%%%%%%%%%%%%%%
\begin{figure*}
\includegraphics[width=\textwidth]{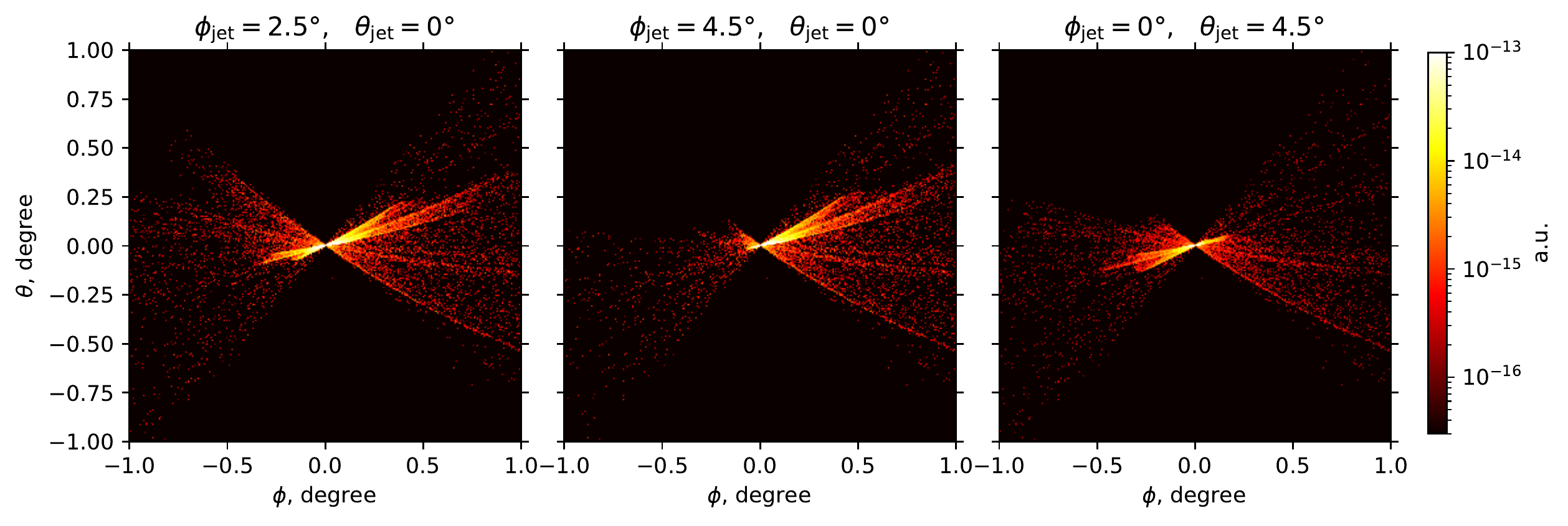}
    \caption{{Images of the extended emission signal around Mkn 501 for different directions of the jet axis with respect to the projected direction of the cosmological magnetic field. Magnetic field strength, energy range and jet opening angle are the same as in Fig.\ref{fig:images}. }
    %\textcolor{red}{Sasha, can you insert the figure with several panels where variations with the orientation of the jet are shown?}
    }
    \label{fig:images1}
\end{figure*}
%%%%%%%%%%%%%%%%%%%%%%%%%%%%%%%%%%%%%%%%%%%%%%%%%%%

The images are superpositions of line-like features that are not perfectly aligned with each other. If the IGMF were perfectly homogeneous, with fixed direction, all these features would be strictly aligned, so that ``narrow-jet-like" extensions would be observable. The appearance of multiple misaligned features reveals the influence of the structure formation on the field. Each feature is produced by the \gr\ beam passing through individual voids along the line of sight. In each void collective motions of the medium induce different deformations of the magnetic field. This results in the changes of directions of deflection of electrons and positrons injected in the voids by the $\gamma\gamma$ pair production. 

Jet-like, rather than halo-like extended emission is generically expected even for short correlation length IGMF (originating from a cosmological phase transition) \cite{2010ApJ...719L.130N}. Thus, the  azimuthal asymmetry of the extended signal is not an indication of the presence of large correlation length IGMF. However, there are two important differences in the properties of the jet-like extended emission in the two alternative cases of phase transition and inflationary magnetic field.  The jet-like extensions are  one-sided in the case of a short correlation length field \cite{2010ApJ...719L.130N}, whereas the  jet-like extensions are two sided for Mkn 421 and Mkn 501 in Fig. \ref{fig:images}. 
%Besides
In addition, in the case of the phase transition IGMF, the position angle of the one-sided jet-like extended emission is determined by the orientation of the jet of the blazar rather than by the direction of the field. It is expected to be random for different sources, whereas in the case of the inflationary large correlation length field the extended sources associated to different blazars have the same position angle. 

If the blazar jet is closely aligned with the line of sight, the two sides of the extended emission are symmetric, they have comparable surface brightness. This symmetry can be broken by the misalignment of the jet with the line of sight. This effect is illustrated in Fig. \ref{fig:images1}. Angles between the blazar jet, the line of sight and the direction of CMF regulate the flux levels of the extended emission components on different sides. 

The orientation of the two-sided extension that is due to the presence of the large correlation length field is, always perpendicular to the direction of magnetic field projected on the sky. If the cosmological IGMF is correlated on the distance scales  larger than hundreds of Megaparsecs, orientations of the two-sided extensions around different sources all over the sky are all expected to be aligned. Measurement of aligned extended emission features around multiple sources can provide an unambiguous evidence for the presence of magnetic field of inflationary origin. In this case position angles of the two-sided extensions provide the measurement of the direction of the primordial field.

%\begin{figure}
%\includegraphics[width=\textwidth]{images_2D_2.pdf}
%    \caption{2D images, $B=10^{-13}$ G}
%    \label{fig:images}
%\end{figure}

%%%%%%%%%%%%%%%%%%%%%%%%%%%%%%%
\begin{figure*}
\includegraphics[width=\textwidth]{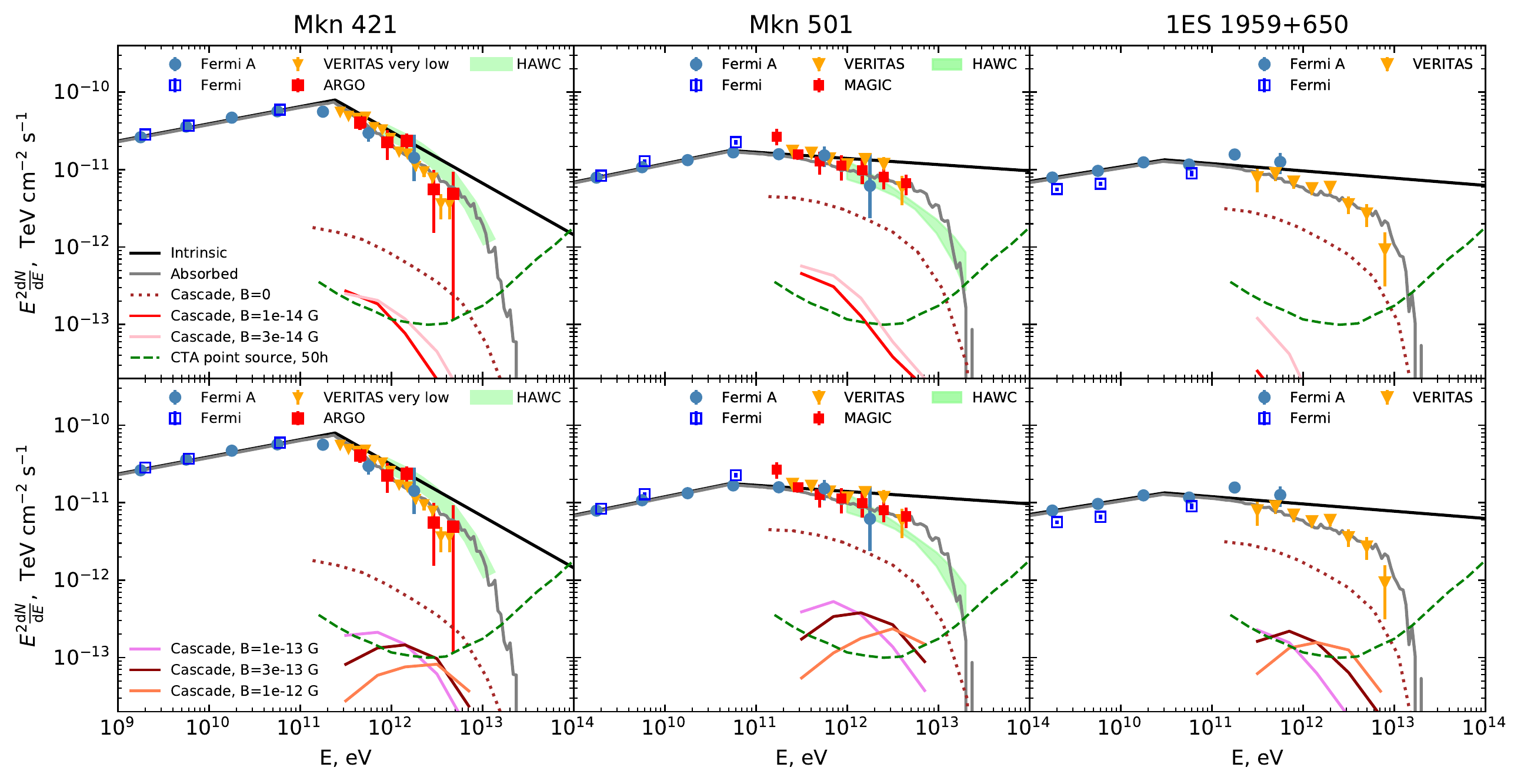}
    \caption{Spectra of intrinsic point source and extended emission for different assumed magnetic field strengths upper panels: $0$, $10^{-14}$ and $3\times 10^{-14}$~G, lower panels: $10^{-13}, 3\times 10^{-13}$ and $10^{-12}$~G), for the three brightest sources in the sample. Point source spectra from Fermi LAT, MAGIC \citep{Mkn501_2009all}, VERITAS \citep{Mkn501_2009all} ARGO \citep{2012ApJ...758....2B} and HAWC \citep{2019ICRC...36..654C}.
    Total flux from the cascade in case $B=0$ is shown with  dotted line. Extended emission outside of the PSF of point source is shown with colored lines for magnetic fields between $10^{-14}$ G and $10^{-12}$ G. Dashed green line show sensitivity of CTA to point sources.
    %\textcolor{red}{would be nice to have legend. Not sure, how did you choose the slope of the green solid line? May be comment on this in the text or update caption. Also add references to the data points. May be show several lines, because we do not know what is the angular width of the wedges that you choose for the measurement of the extended flux... I think you can take the whole extended signal, don't skip the region closest to the point source: we can still measure it, using the PSF measurement in the direction perpendicular to the extension. This will boost a bit the extended source fluxes, so that they would rise above the green solid lines. }
    } 
  
    \label{fig:spec}
\end{figure*}
%%%%%%%%%%%%%%%%%%%%%%%%%%%%%%%

Fig. \ref{fig:spec} shows spectra of the point and extended source emission for different assumptions about the magnetic field strength. For each source we choose the model of the intrinsic spectrum in such a way that it fits the spectral measurements of the "low state" of the source and/or the time-averaged spectral measurements on multi-year time scales.   For all sources we consider two alternative Fermi spectral data: the spectral measurements reported in the Fermi source catalog \cite{fermi-cat} and the spectra extracted usng the aperture photometry approach for the full 13-year exposure of Fermi/LAT up to July 2021. This allows us to extend the Fermi catalog measurements up to the 3 TeV energy\footnote{The ``analysis caveats" page of Fermi Science Support Center, \url{https://fermi.gsfc.nasa.gov/ssc/data/analysis/caveats.html} recommends the use of Fermi/LAT spectra up to 1 TeV, because of the lack of validation of the Instrument Response Functions (IRFs) in the 1-3 TeV energy range. Our comparison of Fermi/LAT and Cherenkov telescope spectra of bright blazars shown in Fig. \ref{fig:spec} shows that the Fermi/LAT measurements in this energy range agree with those of other \gr\ telescopes. }. For the aperture photometry, we extract the source counts: $C_{0.5}$, $C_{1.0}$, from circles of the radius 0.5 and 1 degree using the \textit{gtsselecte-gtmktime-gtbin} chain. We then calculate the exposure (product of the energy-dependent effective area $A$ and exposure time $T$, using the \textit{gtexposure} tool twice: first without the aperture correction, to calculate $AT$, and then with the aperture correction that takes into account the fraction $PSF(\theta)$ of the point spread function contained in the circles of different radii $\theta$: $AT_{0.5,apcorr}=AT\cdot PSF(0.5^\circ)$, $AT_{1.0,apcorr}=AT\cdot PSF(1.0^\circ)$. This allows us to calculate $PSF(0.5^\circ)$ and $PSF(1^\circ)$ that enter a system of equations that determines $C_{0.5}$ and $C_{1.0}$ for the source flux $S$ (measured in counts/(cm$^2$s)) and background flux $B$ (in counts/(cm$^2$sr)):
\begin{eqnarray}
    &&C_{0.5}=AT\left(S\cdot PSF(0.5^\circ)+B\cdot \pi (0.5^\circ)^2\right)\nonumber\\
    &&C_{1.0}=AT\left(S\cdot PSF(1.0^\circ)+B\cdot \pi (1.0^\circ)^2\right).
\end{eqnarray}
Solving this system of equations for $S$ and $B$ we determine the source and background fluxes in each energy bin. The high energy part of Fermi/LAT spectral measurements overlaps well with the energy range of Cherenkov telescope and HAWC measurements. This provides a possibility of cross-calibration of measurements with different instruments. 

For Mkn 501 we used MAGIC and VERITAS analyses from 2009 \citep{Mkn501_2009all}, and long term observations from  ARGO \citep{2012ApJ...758....2B} and HAWC \citep{2019ICRC...36..654C}. 
For Mkn 421 we used long term observations from  ARGO \citep{ARGO-YBJ:2015qiq} and HAWC \citep{2019ICRC...36..654C} and very low state from VERITAS measurement \citep{Acciari:2011kj}.

{We extracted the extended source fluxes from wedges of the angular width $0.3^\circ$ that contain the signal. From the wedges, we excluded the regions in which the primary source emission dominates, namely, we did not take into account those regions of the wedges that are located at a distance from the centers less than the angular resolution of the CTA.}  From Fig. \ref{fig:images} one can judge that the extended emission is dominated by the flux in a much narrower wedge, with opening angle of just a few degrees. 

This gives a conservative estimate of the detectable extended source flux. In principle, stronger signal can be extracted using the azimuthal asymmetry of the extended emission. If the shape of the point spread function is known, the extended emission can be detected also from within the extent of the point spread function. Systematic uncertainty of the knowledge of the point spread function limits the sensitivity of Cherenkov telescopes for detection of extended emission  \cite{2011A&A...526A..90N,2018APh....98....1D}. This uncertainty can be mitigated by the direct measurement of the point spread function in the direction perpendicular to the direction of extended emission. This is possible even if the source line of sight is nearly aligned with the field direction, as illustrated by the example of 1ES 1959+650, shown in the right panel of Fig. \ref{fig:images}.

The linear, one-dimensional, rather than circular, two-dimensional shape of the extended source also provides an improvement of sensitivity for another reason. The sensitivity for extended sources typically worsens with the increase of the solid angle spanned by the extended source on the sky, because of the increase of the background on top of which the signal is detected. Concentration of the extended signal in a narrow wedge reduces the solid angle and hence increases the signal-to-noise ratio.

All the sources in our sample have steep spectra in the 10 TeV range. This diminishes the power of the secondary emission observable in the 0.1-1~TeV range. The total flux that would be available for detection (as a contribution to the point source flux) in the absence of an IGMF is shown by the brown dotted lines in Fig. \ref{fig:spec}. The IGMF deflects electrons and positrons away from the line of sight, so that the secondary \gr s produced by the inverse Compton scattering of the Cosmic Microwave Background becomes unobservable. This effect is stronger at lower energies. This explains the suppression of the extended flux (that we collect from the wedge-shaped regions containing the secondary flux) at the energies below 100~GeV. If the magnetic field is too weak to deflect electrons and positrons, the secondary flux just contributes to the point source flux and is also undetectable. This explains the suppression of the secondary signal at higher energies. Both low- and high-energy suppression depends on the strength of magnetic field. 

Fig. \ref{fig:spec} shows a comparison of the expected secondary flux levels with the sensitivity of the CTA telescopes for detection of extended sources. This comparison shows how challenging the search for the two-sided jet-like extensions  might be. The model predictions for the extended signal are at the limit of sensitivity for the 50~hr exposure of each of the three brightest sources in our source sample. However, use of possible improvements of the method with "in-situ" measurement of the telescope point spread function and careful choice of the wedge for the extended signal measurements may improve the sensitivity. Otherwise, much longer exposure of several hundred hours (instead of 50 hr considered here) can also boost signal-to-noise ratio and make the correlated extended emission signal measureable in all the three brightest nearby blazars considered above.

% \begin{figure}
% \includegraphics[width=\linewidth]{Mrk501_spec.png}
%     \caption{Spectrum for Mkn  501 from Fermi LAT, MAGIC \citep{Mkn501_2009all}, VERITAS \citep{Mkn501_2009all} ARGO \citep{2012ApJ...758....2B} and HAWC \citep{2019ICRC...36..654C}. }
%     \label{fig:spec_mkn501}
% \end{figure}

%\begin{figure}
%\includegraphics[width=1\linewidth]{ES2344_spec.png}
%    \caption{Same as Fig. \ref{fig:spec_mkn501} for $QSO 2344+514$. }
%    \label{fig:spec_mkn501}
%\end{figure}
% \begin{figure}
% \includegraphics[width=1\linewidth]{ES1959_spec.png}
%     \caption{Same as Fig. \ref{fig:spec_mkn501} for $1ES 1959+650$.}
%     \label{fig:spec_mkn501}
% \end{figure}
%\begin{figure}
%\includegraphics[width=1\linewidth]{image_2D_1ES1959.png}
%    \caption{2D image 1ES1959 }
%    \label{fig:spec_mkn501}
%\end{figure}

\section{Conclusions}

In this paper we studied a possibility of detection of primordial magnetic field from inflation  \cite{1992ApJ...391L...1R,Durrer:2013pga} with gamma-ray telescopes. Such field can be coherent on cosmological scales and induce wedge-like extended emission around nearby blazars, with aligned wedge orientations in multiple sources across the sky.

 %Gamma-ray emission from TeV sources generates electromagnetic cascades. Secondary electrons and positrons are deflected by primordial fields in the voids of large scale structure before they up-scatter CMB photons in direction of observer at Earth. As result extended emission around point TeV sources can be detected by their shape, which depends on properties of primordial magnetic field. 
 
This alignment can be used to distinguish the establish the inflationary origin of IGMF, because it is not expected if the IGMF originates from cosmological phase transitions. For small coherence scale IGMF, the wedge like appearance of the extended emission is also generically expected, but the position angles of wedged extended emission around different sources would not be correlated, because it is determined by the orientation of the jets in the primary blazar source. 

The morphology of the secondary emission depends on the jet orientation also for the inflationary magnetic field (Fig. \ref{fig:images1}). However, the jet orientation does not affect the position angle of the extended source. 
%In this paper we studied 7 nearest blazars located within 230 Mpc  around the Earth, for which LSS along the line of sight is known.
%Among those sources Mkn 421, Mkn 501 and 1ES 1959+650 are the sources with best possible discovery potential of IGMF, see Fig. \ref{fig:spec}. 
%In future it will be possible to study line of sight to more distant sources, increasing the number of sources contributing to available for the study of CMF generated by inflation.

\begin{acknowledgements}
The work of A.N., G.L., M.R. and D.S.  has been supported in part by the French National Research Agency (ANR) grant ANR-19-CE31-0020, work of A.K. was supported in part by Russian Science Foundation grant 20-42-09010. A.K.'s stay in the APC laboratory was provided by the  ``Vernadsky'' scholarship of the French embassy in Russia. This work has made use of the Infinity Cluster hosted by Institut d'Astrophysique de Paris. We thank St\'ephane Rouberol for running this cluster smoothly for us. This work has been done within the Aquila Consortium (\url{https://www.aquila-consortium.org}).

\end{acknowledgements}
\bibliography{IGMF_inflation}
\end{document}